\magnification \magstep1
\raggedbottom
\openup 1\jot
\voffset6truemm
\def\cstok#1{\leavevmode\thinspace\hbox{\vrule\vtop{\vbox{\hrule\kern1pt
\hbox{\vphantom{\tt/}\thinspace{\tt#1}\thinspace}}
\kern1pt\hrule}\vrule}\thinspace}
\centerline {\bf Boundary-value problems for the}
\centerline {\bf squared Laplace operator}
\vskip 1cm
\leftline {Giampiero Esposito$^{1,2}$}
\vskip 0.3cm
\noindent
${ }^{1}${\it Istituto Nazionale di Fisica Nucleare, Sezione di Napoli,
Mostra d'Oltremare Padiglione 20, 80125 Napoli, Italy}
\vskip 0.3cm
\noindent
${ }^{2}${\it Universit\`a di Napoli Federico II, Dipartimento
di Scienze Fisiche, Complesso Universitario di Monte S. Angelo,
Via Cintia, Edificio G, 80126 Napoli, Italy}
\vskip 1cm
\noindent
{\bf Summary}. - The squared Laplace operator acting on symmetric
rank-two tensor fields is studied on a (flat) Riemannian 
manifold with smooth boundary. Symmetry of this fourth-order
elliptic operator is obtained provided that such tensor fields
and their first (or second) normal derivatives are set to zero
at the boundary. Strong ellipticity of the resulting boundary-value
problems is also proved. Mixed boundary conditions are eventually
studied which involve complementary projectors and tangential
differential operators. In such a case, strong ellipticity is
guaranteed if a pair of matrices are non-degenerate.
These results find application to the
analysis of quantum field theories on manifolds with boundary.
\vskip 100cm
\leftline {\bf 1. - Introduction}
\vskip 0.3cm
Mathematicians and physicists have become familiar, along the 
years, with the key role played by the operators of Dirac and
Laplace type in the investigation of elliptic operators on 
Riemannian manifolds. For example, it is by now well known that
the symbol of the Dirac operator is a generator of all elliptic
symbols on (closed) Riemannian manifolds [1], and deep results 
in index theory have been found by looking at non-local boundary
conditions for operators of Dirac type [1,2]. More recently, local
supersymmetry and quantum supergravity have led to the consideration
of local boundary conditions for the Dirac operator [1,3--5], 
while theoretical models relevant for quantum chromodynamics rely
on mathematical structures weaker than the standard spin-structures
where, again, a Dirac operator is found to be quite essential [1,6].
This analysis leads, in turn, to a deeper understanding of the geometry
and topology of four-manifolds [6--8]. The operators of Laplace
type, on the other hand, occur naturally in the consideration of
quantized gauge theories (Abelian and non-Abelian) and Euclidean 
quantum gravity [5,9], and many efforts are devoted, within that
framework, to the evaluation of the one-loop semiclassical 
approximation in quantum field theory, with the help of heat-kernel
and $\zeta$-function methods [4,5,10,11].

Nevertheless, there is still room left for the analysis of many other
classes of differential operators on manifolds. In particular, we are
here concerned with the so-called conformally covariant 
operators [12,13]. They arise in the course of studying the 
behaviour of field theories and differential geometric objects under
conformal rescalings of the background metric $g$. More
precisely, if the conformal rescaling of $g$ is written in the form
$g_{\omega}=e^{2 \omega}g$, a conformally covariant operator 
$Q$ satisfies, by definition, the transformation property
$$
Q_{\omega}=e^{-(m+4)\omega /2} \; Q(\omega=0)
\; e^{(m-4)\omega /2} ,
\eqno (1.1)
$$
where $m$ is the dimension of the background Riemannian geometry.
On compact Riemannian manifolds without boundary, the recent
investigations have focused on the structure of anomalies [14]
and on the application to the effective-action formalism [12,13].
For yet other classes of higher order differential operators, 
obtained by composition of operators of Laplace type, much insight
has been gained into the structure of the associated Green 
functions [15,16]. Even more recently, however, fourth-order
elliptic operators have been studied on Riemannian four-manifolds
with boundary, motivated by the path-integral quantization 
programme in gauges which are invariant under conformal 
rescalings [1,17-19]. Of course, even independently of the
particular physical motivation, once that such operators have been
studied for a first time in the presence of boundaries, the analysis
of a broader class of examples appears both natural and desirable, 
to improve the understanding of some key features never studied
before. For this purpose, bearing in mind all motivations mentioned
so far, but leaving aside, for the time being, the immediate physical
applications, we here consider the squared Laplace operator on flat
Riemannian four-manifolds with smooth boundary, with its action on
symmetric rank-two tensor fields on $(M,g)$:
$$
\cstok{\ }^{2} \equiv g^{ab}g^{cd}\nabla_{a}\nabla_{b}
\nabla_{c}\nabla_{d},
\eqno (1.2)
$$
$$
\cstok{\ }^{2}: C^{\infty}(V,M) \rightarrow 
C^{\infty}(V,M).
\eqno (1.3)
$$
With our notation, $\nabla$ is the connection on the vector bundle,
$V$, of symmetric rank-two tensor fields on $M$. On the bundle $V$,
we consider the particular DeWitt super-metric given by
$$
E^{ab \; cd} \equiv {1\over 2} \Bigr(g^{ac}g^{bd}
+g^{ad}g^{bc}\Bigr) +\alpha g^{ab}g^{cd} ,
\eqno (1.4)
$$
with inverse
$$
E_{ab \; cd}^{-1}={1\over 2}\Bigr(g_{ac}g_{bd}+g_{ad}g_{bc}\Bigr)
-{\alpha \over (1+\alpha m)} g_{ab}g_{cd} .
\eqno (1.5)
$$
Since $M$ is taken to be $m$-dimensional, 
with $m \not = 2$, our form of the 
DeWitt super-metric is non-singular, and is positive-definite 
for all $\alpha > -{1\over m}$. 

The step leading to the definition (1.2) is simple but non-trivial,
and hence deserves further comments. Indeed, in Euclidean 
quantum gravity at one loop, it is necessary to add to the
Einstein--Hilbert action a gauge-averaging term, to ensure that
the resulting gauge-field operator $P_{ab}^{\; \; \; cd}$ admits a
Green function and has a non-degenerate leading symbol [5,9].
In particular, the de Donder gauge-averaging term makes it possible
to turn $P_{ab}^{\; \; \; cd}$ 
into an operator of Laplace type (also called
minimal), and in flat space one finds
$$
P_{ab}^{\; \; \; cd}h_{cd}=-\cstok{\ }h_{ab} ,
\eqno (1.6)
$$
where
$$
\cstok{\ } \equiv g^{cd}\nabla_{c}\nabla_{d}
=\nabla^{e}\nabla_{e} .
\eqno (1.7)
$$
Thus, the proof of symmetry of $P_{ab}^{\; \; \; cd}$ is reduced to proving 
that the operator $\cstok{\ }$ defined in (1.7) is symmetric. 
In a flat Riemannian background, an analysis along similar lines
leads to the consideration of the operator (1.2) starting from the
composition of the operator $P_{ab}^{\; \; \; cd}$ with itself, i.e.
$$
P_{ab}^{\; \; \; cd} \; P_{cd}^{\; \; \; ef} 
\; h_{ef}=\cstok{\ }^{2}h_{ab} ,
\eqno (1.8)
$$
and hence the object of interest is actually the $\cstok{\ }^{2}$
operator (although we are leaving aside the problem of deriving the
full gauge-field operator on manifolds with boundary for 
curvature-squared theories of gravity).

Section {\bf 2} studies the inner product on the space of smooth 
and symmetric rank-two tensor fields on $M$, and performs the
integrations by parts which are necessary to understand whether
$\cstok{\ }^{2}$ can be realized as a symmetric operator with
suitable boundary conditions. A first set of boundary conditions,
which make it possible to achieve the desired symmetry, are
obtained in sect. {\bf 3}. A second set of admissible boundary conditions
in instead derived in sect. {\bf 4}. 
Section {\bf 5} proves that the resulting
boundary-value problems are strongly elliptic, and extends such
a proof to the case of mixed boundary conditions for the squared
Laplacian. Concluding remarks are presented in sect. {\bf 6}.
\vskip 0.3cm
\leftline {\bf 2. - Inner product and integration by parts}
\vskip 0.3cm 
In our paper, we use the following definition of inner product:
$$
(\eta,h) \equiv \int_{M} \eta_{ab} E^{ab \; cd} h_{cd}
\sqrt{g} \; d^{m}x,
\eqno (2.1)
$$
where $E^{ab \; cd}$ is the DeWitt super-metric (1.4), and 
$\eta_{ab},h_{ab}$ are any two sections of the vector bundle of
smooth, symmetric rank-two tensor fields over the $m$-dimensional
Riemannian manifold $(M,g)$. The definition (2.1) gives rise
to an inner product if $\alpha+{1\over m} >0$,
because by virtue of (1.4) one finds
$$
(\eta,\eta)=\int_{M}\left[\eta_{ab}\eta^{ab} +\alpha
\Bigr(\eta_{\; a}^{a}\Bigr)^{2}\right] \sqrt{g} \; d^{m}x,
\eqno (2.2)
$$
and, after using the general decomposition of $\eta_{ab}$, the
coefficient of $\Bigr(\eta_{\; a}^{a}\Bigr)^{2}$ in the integrand
(2.2) is found to be $\alpha+{1\over m}$. Hereafter we shall 
therefore restrict $\alpha$ to be greater than $-{1\over m}$.

If $\cstok{\ }^{2}$ were a symmetric operator, the scalar products
$\Bigr(\eta,\cstok{\ }^{2}h \Bigr)$ and 
$\Bigr(\cstok{\ }^{2}\eta,h \Bigr)$ should be equal for all 
$\eta,h$ in the domain of $\cstok{\ }^{2}$. In our paper, 
the above difference reads (hereafter, we assume $M$ is
four-dimensional)
$$
\eqalignno{
\; & I(\eta,h) \equiv \Bigr(\eta,\cstok{\ }^{2}h \Bigr)
-\Bigr(\cstok{\ }^{2}\eta, h \Bigr) \cr
&=\int_{M}\biggr[\eta_{ab}\nabla^{c}\nabla_{c}
\nabla^{d}\nabla_{d}\Bigr(h^{ab}+\alpha g^{ab}h \Bigr) \cr
&-h_{ab}\nabla^{c}\nabla_{c}\nabla^{d}\nabla_{d}
\Bigr(\eta^{ab}+\alpha g^{ab}\eta \Bigr)\biggr]
\sqrt{g} \; d^{4}x \cr
&=\int_{\partial M}n^{p}\biggr[\eta_{00}\nabla_{p}
\cstok{\ }\Bigr(h^{00} +\alpha h \Bigr)
-h_{00}\nabla_{p}\cstok{\ }\Bigr(\eta^{00} +\alpha\eta\Bigr) \cr
&+2 \eta_{0i}\nabla_{p}\cstok{\ }h^{0i}
-2h_{0i}\nabla_{p}\cstok{\ }\eta^{0i} \cr
&+\eta_{ij}\nabla_{p}\cstok{\ }\Bigr(h^{ij} +\alpha
g^{ij}h \Bigr)-h_{ij}\nabla_{p}\cstok{\ }
\Bigr(\eta^{ij} +\alpha g^{ij}\eta \Bigr)
\biggr]\sqrt{\gamma} \; d^{3}x,
&(2.3)\cr}
$$
where we have assumed that $g^{00}=1$, $g^{0i}=0$, 
$n^{p}=(1,0,0,0)$
is the unit normal to the boundary, and $\gamma$ is the determinant
of the induced three-metric on $\partial M$. Of course, $h$ and
$\eta$ in the integrand (2.3) 
denote the traces $h \equiv g^{ab}h_{ab}$, 
$\eta \equiv g^{ab}\eta_{ab}$. It is now convenient to define, 
for $\varphi_{ab}=\eta_{ab},h_{ab}$,
$$
F^{00}(\varphi) \equiv \varphi^{00} +\alpha g^{00}{\varphi}
=\varphi^{00} +\alpha \varphi ,
\eqno (2.4)
$$
$$
F^{0i}(\varphi) \equiv \varphi^{0i} +\alpha g^{0i}\varphi
=\varphi^{0i} ,
\eqno (2.5)
$$
$$
F^{ij}(\varphi) \equiv \varphi^{ij}
+\alpha g^{ij}\varphi,
\eqno (2.6)
$$
so that Eq. (2.3) is expressed in the more convenient form
$$ 
\eqalignno{
\; & \Bigr(\eta,\cstok{\ }^{2}h \Bigr)
- \Bigr(\cstok{\ }^{2}\eta, h \Bigr) \cr
&=\int_{\partial M}n^{p} \Bigr[\eta_{00}\cstok{\ }
\nabla_{p}F^{00}(h)-h_{00}\cstok{\ } \nabla_{p}F^{00}(\eta)
\Bigr]\sqrt{\gamma} \; d^{3}x \cr
&+2 \int_{\partial M}n^{p}\Bigr[\eta_{0i}\cstok{\ }
\nabla_{p}F^{0i}(h)-h_{0i}\cstok{\ }\nabla_{p} F^{0i}(\eta)
\Bigr]\sqrt{\gamma} \; d^{3}x \cr
&+\int_{\partial M}n^{p}\Bigr[\eta_{ij}\cstok{\ }
\nabla_{p}F^{ij}(h)-h_{ij}\cstok{\ }\nabla_{p}F^{ij}(\eta)
\Bigr]\sqrt{\gamma} \; d^{3}x,
&(2.7)\cr}
$$
where we have used the well known commutation property of
covariant derivatives in a background whose Riemann curvature
vanishes. Our aim is now to use integration by parts so as to
remove the third-order derivatives occurring in Eq. (2.7).
For this purpose, we use repeatedly the Leibniz rule. The
analysis of the first boundary integral in Eq. (2.7) suggests
considering (hereafter, $K_{ij}$ is the extrinsic-curvature
tensor of the boundary, with trace $K$)
$$
\eqalignno{
\nabla_{p}\Bigr(n^{p}\eta_{00}\cstok{\ }F^{00}(h)\Bigr)
&=K \eta_{00} \cstok{\ } F^{00}(h)+n^{p}(\nabla_{p}\eta_{00})
\cstok{\ }F^{00}(h) \cr
&+n^{p}\eta_{00}\cstok{\ } \nabla_{p} F^{00}(h),
&(2.8)\cr}
$$
jointly with
$$ 
\eqalignno{
\; & \nabla_{c} \Bigr(n^{p}(\nabla_{p}\eta_{00})
\nabla^{c}F^{00}(h)\Bigr)=K_{c}^{\; p}(\nabla_{p}\eta_{00})
\nabla^{c}F^{00}(h) \cr
&+n^{p}(\nabla_{c}\nabla_{p}\eta_{00})\nabla^{c}F^{00}(h)
+n^{p}(\nabla_{p}\eta_{00})\cstok{\ }F^{00}(h).
&(2.9)\cr}
$$
Moreover, the analysis of the second boundary integral in Eq.
(2.7) suggests using the identity
$$ 
\eqalignno{
\nabla_{p} \Bigr(n^{p}\eta_{0i}\cstok{\ }F^{0i}(h)\Bigr)
&=K \eta_{0i} \cstok{\ } F^{0i}(h)
+n^{p}(\nabla_{p}\eta_{0i})\cstok{\ }F^{0i}(h) \cr
&+n^{p}\eta_{0i} \cstok{\ } \nabla_{p}F^{0i}(h),
&(2.10)\cr}
$$
and
$$
\eqalignno{
\; & \nabla_{c}\Bigr(n^{p}(\nabla_{p}\eta_{0i})
\nabla^{c}F^{0i}(h)\Bigr)=K_{c}^{\; p}(\nabla_{p}\eta_{0i})
\nabla^{c}F^{0i}(h) \cr
&+n^{p}(\nabla_{c}\nabla_{p}\eta_{0i})\nabla^{c}F^{0i}(h)
+n^{p}(\nabla_{p}\eta_{0i})\cstok{\ }F^{0i}(h).
&(2.11)\cr}
$$
Furthermore, the consideration of the third boundary integral
in Eq. (2.7) makes it convenient to use the identities
$$ \eqalignno{
\nabla_{p}\Bigr(n^{p}\eta_{ij}\cstok{\ }F^{ij}(h)\Bigr)&=
K \eta_{ij}\cstok{\ }F^{ij}(h)+n^{p}(\nabla_{p}\eta_{ij})
\cstok{\ }F^{ij}(h) \cr
&+n^{p}\eta_{ij}\cstok{\ }\nabla_{p}F^{ij}(h),
&(2.12)\cr}
$$
$$ \eqalignno{
\; & \nabla_{c}\Bigr(n^{p}(\nabla_{p}\eta_{ij})\nabla^{c}
F^{ij}(h)\Bigr)=K_{c}^{\; p}(\nabla_{p}\eta_{ij})
\nabla^{c}F^{ij}(h) \cr
&+n^{p}(\nabla_{c}\nabla_{p}\eta_{ij})\nabla^{c}F^{ij}(h)
+n^{p}(\nabla_{p}\eta_{ij})\cstok{\ }F^{ij}(h).
&(2.13)\cr}
$$
If the boundary of $M$ is smooth, one has $\partial \partial M
=\emptyset$. Moreover, the integral vanishes over zero-measure
sets. It is hence possible to use Eqs. (2.8)--(2.13) and the
Stokes theorem to re-express Eq. (2.7) in the form
$$
\eqalignno{
\; & \Bigr(\eta,\cstok{\ }^{2}h \Bigr)
-\Bigr(\cstok{\ }^{2}\eta,h \Bigr) \cr
&=\int_{\partial M}\biggr[-K \Bigr(\eta_{00}\cstok{\ }
F^{00}(h)-h_{00}\cstok{\ }F^{00}(\eta)\Bigr)\cr
&+K_{i}^{\; j}\Bigr((\nabla_{j}\eta_{00})\nabla^{i}F^{00}(h)
-(\nabla_{j}h_{00})\nabla^{i}F^{00}(\eta)\Bigr) \cr
&+n^{p}\Bigr((\nabla_{c}\nabla_{p}\eta_{00})
\nabla^{c}F^{00}(h)-(\nabla_{c}\nabla_{p}h_{00})
\nabla^{c}F^{00}(\eta)\Bigr) \cr
&-2K \Bigr(\eta_{0i}\cstok{\ }F^{0i}(h)
-h_{0i}\cstok{\ }F^{0i}(\eta)\Bigr) \cr
&+2K_{l}^{\; j}\Bigr((\nabla_{j}\eta_{0i})
\nabla^{l}F^{0i}(h)-(\nabla_{j}h_{0i})\nabla^{l}F^{0i}(\eta)
\Bigr) \cr
&+2n^{p}\Bigr((\nabla_{c}\nabla_{p}\eta_{0i})
\nabla^{c}F^{0i}(h) 
-(\nabla_{c}\nabla_{p}h_{0i})\nabla^{c}F^{0i}(\eta)\Bigr) \cr
&-K \Bigr(\eta_{ij}\cstok{\ }F^{ij}(h)
-h_{ij}\cstok{\ }F^{ij}(\eta) \Bigr) \cr
&+K_{l}^{\; r}\Bigr((\nabla_{r}\eta_{ij})
\nabla^{l}F^{ij}(h)-(\nabla_{r}h_{ij})
\nabla^{l}F^{ij}(\eta)\Bigr) \cr
&+n^{p}\Bigr((\nabla_{c}\nabla_{p}\eta_{ij})
\nabla^{c}F^{ij}(h)-(\nabla_{c}\nabla_{p}h_{ij})
\nabla^{c}F^{ij}(\eta)\Bigr)
\biggr] \sqrt{\gamma} \; d^{3}x . 
&(2.14)\cr}
$$
\vskip 0.3cm
\leftline {\bf 3. - Boundary conditions: first option}
\vskip 0.3cm
We know from the work of ref. [19], which studied the action of
the squared Laplace operator on scalar functions 
$f \in C^{\infty}(M)$, that one can impose boundary conditions
where $f$ and its normal derivative vanish at the boundary. This
``doubling" of the boundary conditions, with respect to the
analysis of the Laplacian, is clearly understood if one thinks 
of the eigenvalue problem. In other words, given a spectral
resolution of a fourth-order elliptic operator, one deals with
fourth-order eigenvalue equations which admit four linearly
independent integrals. If one were to fix just the eigenfunctions
at the boundary, one would not get enough equations to determine
the coefficients of linear combination in the equation
$$
f_{\lambda}(x)=C_{1,\lambda}f_{1,\lambda}(x)
+C_{2,\lambda}f_{2,\lambda}(x)
+C_{3,\lambda}f_{3,\lambda}(x)
+C_{4,\lambda}f_{4,\lambda}(x)
$$
for the eigenfunction belonging to the eigenvalue $\lambda$. 
At a deeper level, as shown in ref. [19], one has to integrate
by parts to prove that suitable boundary conditions exist for 
which the operator $\cstok{\ }^{2}$ is (essentially) self-adjoint
(see Eq. (6.1)).

In our problem, the technical details are more elaborated, since
we study the action of $\cstok{\ }^{2}$ on smooth, symmetric
rank-two tensor fields on a flat Riemannian manifold with smooth
boundary (e.g. the Euclidean four-ball), but Eq. (2.14) can be 
used to derive all admissible sets of boundary conditions. First,
we consider a scheme where half of the boundary conditions 
consist of requiring that all components of $\eta_{ab}$ and 
$h_{ab}$, both spatial and normal, should vanish at the boundary.
Thus, we require that
$$
\Bigr[\eta_{00}\Bigr]_{\partial M}
=\Bigr[h_{00}\Bigr]_{\partial M}=0,
\eqno (3.1)
$$
$$
\Bigr[\eta_{0i}\Bigr]_{\partial M}
=\Bigr[h_{0i}\Bigr]_{\partial M}=0,
\eqno (3.2)
$$
$$
\Bigr[\eta_{ij}\Bigr]_{\partial M}
=\Bigr[h_{ij}\Bigr]_{\partial M}=0.
\eqno (3.3)
$$
By virtue of (3.1)--(3.3) one then finds that the
following covariant derivatives vanish at the boundary of $M$
(see (3.20) and (3.21)):
$$
\Bigr[\nabla_{j}\eta_{00}\Bigr]_{\partial M}
=\Bigr[\nabla_{j}h_{00}\Bigr]_{\partial M}=0,
\eqno (3.4)
$$
$$
\Bigr[\nabla_{j}\eta_{0i}\Bigr]_{\partial M}
=\Bigr[\nabla_{j}h_{0i}\Bigr]_{\partial M}=0,
\eqno (3.5)
$$
$$
\Bigr[\nabla_{k}\eta_{ij}\Bigr]_{\partial M}
=\Bigr[\nabla_{k}h_{ij}\Bigr]_{\partial M}=0.
\eqno (3.6)
$$
At this stage, only the third, sixth 
and ninth line give non-vanishing
contributions to the integrand in Eq. (2.14). It is hence 
appropriate to write them explicitly, bearing in mind, from
sect. {\bf 2}, that the normal to the boundary takes the form
$n^{p}=(1,0,0,0)$. In other words, one has (with $c,p$ ranging
from 0 through 3, and $i,k$ ranging from 1 through 3)
$$
n^{p}(\nabla_{c}\nabla_{p}\eta_{00})\nabla^{c}F^{00}(h)
=(\nabla_{0}\nabla_{0}\eta_{00})\nabla^{0}F^{00}(h)
+(\nabla_{k}\nabla_{0}\eta_{00})\nabla^{k}F^{00}(h),
\eqno (3.7)
$$
$$
n^{p}(\nabla_{c}\nabla_{p}\eta_{0i})\nabla^{c}F^{0i}(h)
=(\nabla_{0}\nabla_{0}\eta_{0i})\nabla^{0}F^{0i}(h)
+(\nabla_{k}\nabla_{0}\eta_{0i})\nabla^{k}F^{0i}(h),
\eqno (3.8)
$$
$$
n^{p}(\nabla_{c}\nabla_{p}\eta_{ij})\nabla^{c}F^{ij}(h)
=(\nabla_{0}\nabla_{0}\eta_{ij})\nabla^{0}F^{ij}(h)
+(\nabla_{k}\nabla_{0}\eta_{ij})\nabla^{k}F^{ij}(h),
\eqno (3.9)
$$
and another triple of identities, which also contribute to the 
third, sixth and ninth line of (2.14), and are obtained by interchanging 
the roles of $\eta_{ab}$ and $h_{ab}$ in (3.7)--(3.9). The
right-hand sides of (3.7)--(3.9) are linear combinations of
products of covariant derivatives of 
$\eta_{ab}$ and $h_{ab}$. Thus, it is
sufficient to set to zero at the boundary only one of the two
functions occurring in the product. For example, to avoid setting
to zero on $\partial M$ the second normal derivatives of $\eta_{00}$,
$\eta_{0i}$ and $\eta_{ij}$ 
(and hence of $h_{00}$, $h_{0i}$ and $h_{ij}$ as well) one
can require that
$$
\Bigr[\nabla_{0}\eta_{00}\Bigr]_{\partial M}
=\Bigr[\nabla_{0}h_{00}\Bigr]_{\partial M}=0,
\eqno (3.10)
$$
$$
\Bigr[\nabla_{0}\eta_{0i}\Bigr]_{\partial M}
=\Bigr[\nabla_{0}h_{0i}\Bigr]_{\partial M}=0,
\eqno (3.11)
$$
$$
\Bigr[\nabla_{0}\eta_{ij}\Bigr]_{\partial M}
=\Bigr[\nabla_{0}h_{ij}\Bigr]_{\partial M}=0,
\eqno (3.12)
$$
$$
\Bigr[\nabla^{0}F^{00}(h)\Bigr]_{\partial M}
=\Bigr[\nabla^{0}F^{00}(\eta)\Bigr]_{\partial M}=0,
\eqno (3.13)
$$
$$
\Bigr[\nabla^{0}F^{0i}(h)\Bigr]_{\partial M}
=\Bigr[\nabla^{0}F^{0i}(\eta)\Bigr]_{\partial M}=0,
\eqno (3.14)
$$
$$
\Bigr[\nabla^{0}F^{ij}(h)\Bigr]_{\partial M}
=\Bigr[\nabla^{0}F^{ij}(\eta)\Bigr]_{\partial M}=0.
\eqno (3.15)
$$
Note that we do not get an overdetermined problem, because, by
virtue of (2.4)--(2.6) and (3.10)--(3.12), Eqs. (3.13)--(3.15)
are satisfied. Now it is 
helpful to consider an example, and for this
purpose we choose the Euclidean four-ball, whose metric may be
locally cast in the form
$$
g=d\tau \otimes d\tau +\tau^{2}c_{ij}dx^{i} \otimes dx^{j},
\eqno (3.16)
$$
where the radial coordinate $\tau$ lies in the closed interval
$[0,a]$, with $a$ the radius of the three-sphere boundary, and
$c_{ij}$ is the metric on a unit three-sphere, with local 
coordinates $\left \{ x^{i} \right \}$. One then finds
(Latin indices run here from 1 through 3)
$$
\nabla_{0}h_{00}={\partial h_{00}\over \partial \tau},
\eqno (3.17)
$$
$$
\nabla_{0}h_{0i}={\partial h_{0i}\over \partial \tau}
-{1\over \tau}h_{0i},
\eqno (3.18)
$$
$$
\nabla_{0}h_{ij}={\partial h_{ij}\over \partial \tau}
-{2\over \tau}h_{ij},
\eqno (3.19)
$$
$$
\nabla_{k}h_{00}={\partial h_{00}\over \partial x^{k}}
-{2\over \tau}h_{0k},
\eqno (3.20)
$$
$$
\nabla_{k}h_{0i}={\partial h_{0i}\over \partial x^{k}}
-\Gamma_{\; ki}^{l}h_{l0}-{1\over \tau}h_{ik}
+{1\over \tau}g_{ik}h_{00},
\eqno (3.21)
$$
with $\Gamma$ used to denote the connection coefficients,
and hence
$$
\nabla_{k}\nabla_{0}h_{00}={\partial \over \partial x^{k}}
{\partial h_{00}\over \partial \tau}-{1\over \tau}
\left({\partial h_{00}\over \partial x^{k}}
+2{\partial h_{0k}\over \partial \tau}
-{4\over \tau}h_{0k}\right),
\eqno (3.22)
$$
$$ \eqalignno{
\nabla_{k}\nabla_{0}h_{0i}&={\partial \over \partial x^{k}}
\left({\partial h_{0i}\over \partial \tau}-{1\over \tau}h_{0i}
\right)-{1\over \tau}\left({\partial h_{0i}\over \partial x^{k}}
-\Gamma_{\; ki}^{l}h_{l0}\right) \cr
&-\Gamma_{\; ik}^{l} \left({\partial h_{0l}\over \partial \tau}
-{1\over \tau}h_{0l}\right)+{1\over \tau}g_{ik}
\left({\partial h_{00}\over \partial \tau}
-{1\over \tau}h_{00}\right) \cr
&-{1\over \tau}\left({\partial h_{ik}\over \partial \tau}
-{3\over \tau}h_{ik} \right),
&(3.23)\cr}
$$
$$ \eqalignno{
\nabla_{k}\nabla_{0}h_{ij}&={\partial \over \partial x^{k}}
\left({\partial h_{ij}\over \partial \tau}
-{2\over \tau}h_{ij}\right)-{1\over \tau}\nabla_{k}h_{ij} \cr
&+{2\over \tau}g_{k(i} \left({\partial \over \partial \tau}
-{1\over \tau}\right)h_{j)0} 
-2 \Gamma_{\; k(i}^{l} \left({\partial \over \partial \tau}
-{2\over \tau}\right)h_{j)l} .
&(3.24)\cr}
$$
The joint effect of Eqs. (3.1)--(3.3), (3.7)--(3.12) is then
that the right-hand side of Eq. (2.14) vanishes if the
following boundary conditions hold:
$$
\Bigr[h_{ab}\Bigr]_{\partial M}
=\Bigr[\eta_{ab}\Bigr]_{\partial M}=0 \; \;
\forall a,b=0,1,2,3,
\eqno (3.25)
$$
$$
\left[{\partial h_{ab}\over \partial \tau}\right]_{\partial M}
=\left[{\partial \eta_{ab}\over \partial \tau}\right]_{\partial M}=0
\; \; \forall a,b=0,1,2,3 .
\eqno (3.26a)
$$
One then deals with 10 boundary conditions on $h_{ab},\eta_{ab}$
and 10 boundary conditions on their normal derivatives, bearing 
in mind the symmetry of these rank-two tensor fields. This is a
generalization of the boundary conditions obtained in ref. [19] for
the squared Laplacian acting on smooth functions, and once more 
the number of boundary conditions is exactly doubled, with respect
to the analysis of the Laplacian. Last, a covariant form of Eq.
(3.26a) is easily obtained from (3.10)--(3.12), i.e.
$$
\Bigr[n^{p}\nabla_{p}h_{ab}\Bigr]_{\partial M}
=\Bigr[n^{p}\nabla_{p}\eta_{ab}\Bigr]_{\partial M}=0 \; \; 
\forall a,b=0,1,2,3 .
\eqno (3.26b)
$$
The normal components of $\varphi_{ab}=h_{ab}$ or $\eta_{ab}$
are given by
$$
\varphi_{00}=\varphi_{ab}n^{a}n^{b} ,
\eqno (3.27)
$$
$$
\varphi_{0i}=\varphi_{ai}n^{a} ,
\eqno (3.28)
$$
and the spatial components are obtained, instead, by applying 
a projection operator, i.e.
$$
\varphi_{ij}=\Pi_{ij}^{\; \; \; cd} \; \varphi_{cd} ,
\eqno (3.29)
$$
where, on defining [20]
$$
q_{a}^{\; b} \equiv \delta_{a}^{\; b}-n_{a}n^{b} ,
\eqno (3.30)
$$
one has [20]
$$
\Pi_{ab}^{\; \; \; cd} \equiv q_{(a}^{\; \; c} \;
q_{b)}^{\; \; d} .
\eqno (3.31)
$$
The tensor field $q_{a}^{\; b}$ is the standard projector of tensor
fields over the bounding surface, and is, by construction, 
orthogonal to the unit normal vector, in that
$q_{a}^{\; b}n_{b}=n_{a}-n_{a}=0$.
\vskip 0.3cm
\leftline {\bf 4. - Boundary conditions: second option}
\vskip 0.3cm
After writing down the identities (3.7)--(3.9), one can
also make the alternative choice, according to which
(cf. (3.10)--(3.15))
$$
\Bigr[\nabla_{0}\nabla_{0}\eta_{00}\Bigr]_{\partial M}
=\Bigr[\nabla_{0}\nabla_{0}h_{00}\Bigr]_{\partial M}=0,
\eqno (4.1)
$$
$$
\Bigr[\nabla_{0}\nabla_{0}\eta_{0i}\Bigr]_{\partial M}
=\Bigr[\nabla_{0}\nabla_{0}h_{0i}\Bigr]_{\partial M}=0,
\eqno (4.2)
$$
$$
\Bigr[\nabla_{0}\nabla_{0}\eta_{ij}\Bigr]_{\partial M}
=\Bigr[\nabla_{0}\nabla_{0}h_{ij}\Bigr]_{\partial M}=0,
\eqno (4.3)
$$
$$
\Bigr[\nabla^{k}F^{00}(h)\Bigr]_{\partial M}
=\Bigr[\nabla^{k}F^{00}(\eta)\Bigr]_{\partial M}=0,
\eqno (4.4)
$$
$$
\Bigr[\nabla^{k}F^{0i}(h)\Bigr]_{\partial M}
=\Bigr[\nabla^{k}F^{0i}(\eta)\Bigr]_{\partial M}=0,
\eqno (4.5)
$$
$$
\Bigr[\nabla^{k}F^{ij}(h)\Bigr]_{\partial M}
=\Bigr[\nabla^{k}F^{ij}(\eta)\Bigr]_{\partial M}=0.
\eqno (4.6)
$$
Now the boundary conditions (3.1)--(3.3), jointly with the
definitions (2.4)--(2.6), imply that Eqs. (4.4)--(4.6) are
identically satisfied. In particular, on the Euclidean four-ball,
by virtue of the formulae
$$
\nabla_{0}\nabla_{0}h_{00}={\partial^{2}h_{00}\over 
\partial \tau^{2}},
\eqno (4.7)
$$
$$
\nabla_{0}\nabla_{0}h_{0i}={\partial^{2}h_{0i}\over 
\partial \tau^{2}}-{2\over \tau}
{\partial h_{0i}\over \partial \tau}+{2\over \tau^{2}}
h_{0i},
\eqno (4.8)
$$
$$
\nabla_{0}\nabla_{0}h_{ij}={\partial^{2}h_{ij}\over \partial
\tau^{2}}-{4\over \tau}{\partial h_{ij}\over \partial \tau}
+{6\over \tau^{2}}h_{ij},
\eqno (4.9)
$$
the boundary conditions (3.1)--(3.3) and (4.1)--(4.3) lead to
10 homogeneous Dirichlet conditions, jointly with 10 boundary 
conditions for a linear combination of first- and second-order
partial derivatives, i.e.
$$
\left[{\partial^{2}\eta_{00}\over \partial \tau^{2}}
\right]_{\partial M}=\left[{\partial^{2}h_{00}\over
\partial \tau^{2}}\right]_{\partial M}=0,
\eqno (4.10)
$$
$$
\left[{\partial^{2}\eta_{0i}\over \partial \tau^{2}}
-{2\over \tau}{\partial \eta_{0i}\over \partial \tau}
\right]_{\partial M}
=\left[{\partial^{2}h_{0i}\over \partial \tau^{2}}
-{2\over \tau}{\partial h_{0i}\over \partial \tau}
\right]_{\partial M}=0,
\eqno (4.11)
$$
$$
\left[{\partial^{2}\eta_{ij}\over \partial \tau^{2}}
-{4\over \tau}{\partial \eta_{ij}\over \partial \tau}
\right]_{\partial M}
=\left[{\partial^{2}h_{ij}\over \partial \tau^{2}}
-{4\over \tau}{\partial h_{ij}\over \partial \tau}
\right]_{\partial M}=0.
\eqno (4.12)
$$
Of course, the boundary conditions (4.1)--(4.3) can also be
expressed in a covariant form by a single equation, i.e.
$$
\Bigr[n^{p}n^{q}\nabla_{p}\nabla_{q}\eta_{ab}\Bigr]_{\partial M}
=\Bigr[n^{p}n^{q}\nabla_{p}\nabla_{q}h_{ab}\Bigr]_{\partial M}=0
\; \forall a,b=0,1,2,3 .
\eqno (4.13)
$$
As in sect. {\bf 3}, the number of boundary conditions is doubled with
respect to the analysis of the Laplacian, and Eqs. (3.1)--(3.3)
and (4.1)--(4.3) provide the generalization of yet another set of
boundary conditions derived, in ref. [19], when the action of the
squared Laplacian on smooth functions is considered.
\vskip 1cm
\leftline {\bf 5. - Strong ellipticity}
\vskip 0.3cm
If a differential operator is studied on a Riemannian manifold
$M$ with smooth boundary $\partial M$, the ellipticity is obtained
upon proving that the leading symbol is elliptic in the interior of
$M$, and that a unique solution exists of the eigenvalue equation
for the leading symbol, subject to a decay condition at infinity 
and to suitable boundary conditions. For the squared Laplace operator
defined in Eq. (1.2), the leading symbol
$$
\sigma_{L}\Bigr(\cstok{\ }^{2};x,\xi \Bigr)=|\xi|^{4}I
=g^{\mu \nu}g^{\rho \sigma}\xi_{\mu}\xi_{\nu}\xi_{\rho}
\xi_{\sigma} I,
\eqno (5.1)
$$
with $\xi \in T^{*}(M)$, is clearly elliptic. In fact, for any
Riemannian metric $g$, the leading symbol in (5.1) is 
non-degenerate for $\xi \not = 0$. Moreover, for any 
$\lambda \in {\cal C}-{\Re}_{+}$, one has
$$
{\rm det}\left(\sigma_{L}\Bigr(\cstok{\ }^{2};x,\xi \Bigr)
-\lambda \right)
={\Bigr(|\xi|^{4}-\lambda \Bigr)}^{{\rm dim}(V)} \not = 0.
\eqno (5.2)
$$
The left-hand side of Eq. (5.2) vanishes if and only if both $\xi$
and $\lambda$ vanish. This completes the proof of ellipticity of
$\sigma_{L}(\cstok{\ }^{2})$ in the interior of $M$.

Now we have to define the condition of strong ellipticity, which
involves in a crucial way the boundary conditions. For this purpose,
we consider again the leading symbol $\sigma_{L}\Bigr(\cstok{\ }^{2};
{\hat x},r,\zeta,\omega \Bigr)$, with $\left \{ {\hat x}^{k} \right \}$
local coordinates on $\partial M$, $r$ the normal geodesic distance 
to the boundary, $\zeta_{j}$ as components of a cotangent vector
on the boundary, and $\omega$ a real parameter. We then set $r=0$
and replace $\omega$ by $-i \partial_{r}$, and consider the eigenvalue
equation for $\sigma_{L}$, i.e. [9,10]
$$
\sigma_{L}\Bigr(\cstok{\ }^{2};{\hat x},r=0,\zeta, -i\partial_{r}
\Bigr)\varphi(r)=\lambda \varphi(r).
\eqno (5.3)
$$
Note that, in writing down Eq. (5.3), we have not doubled the
number of arguments in the leading symbol. In the presence of
boundaries, the $m$ local coordinates $x$ are split into $m-1$ 
local coordinates on the boundary, jointly with the normal
geodesic distance (which, of course, vanishes on the boundary).
Similarly, the $m$ coordinates $\xi_{\mu}$ are split into $m-1$
coordinates on $T^{*}({\partial M})$, appropriate for cotangent
vectors, jointly with a real parameter $\omega$. Hence 
one has (with $|\zeta|^{2} \equiv \zeta_{i}\zeta^{i}$)
$$ 
\sigma_{L}\Bigr(\cstok{\ }^{2}; \left \{ {\hat x}^{k}
\right \},r=0, \left \{ \zeta_{j} \right \}, \omega \Bigr)
=|\zeta|^{4}+\omega^{4}+2\omega^{2}|\zeta|^{2}.
$$
Last, to obtain the left-hand side of Eq. (5.3), one has to
replace $\omega$ by $-i \partial_{r}$ as we said before [10],
and this leads to the fourth-order differential operator
$$
\sigma_{L}\Bigr(\cstok{\ }^{2};\left \{ {\hat x}^{k} \right \},
r=0, \left \{ \zeta_{j} \right \}, 
\omega=-i\partial_{r} \Bigr)
={\partial^{4}\over \partial r^{4}}
-2 |\zeta|^{2}{\partial^{2}\over \partial r^{2}}
+|\zeta|^{4}.
$$
Thus, Eq. (5.3) should be viewed as an 
ordinary differential equation involving
the field $\varphi$ which is a smooth section of the vector bundle
$V$ over the product manifold ${\partial M} \times {\Re}_{+}$. By
definition, for a given form of the boundary operator $B$, the
boundary-value problem $\Bigr(\cstok{\ }^{2},B \Bigr)$ is strongly
elliptic with respect to the cone ${\cal C}-{\Re}_{+}$ if there exists a
unique solution of Eq. (5.3) with $\lambda \in {\cal C}-{\Re}_{+}$,
for all $(\zeta,\lambda) \not = (0,0)$, subject to the asymptotic
condition
$$
\lim_{r \to \infty}\varphi(r)=0,
\eqno (5.4)
$$
and to the boundary condition
$$
\sigma_{g}(B)({\hat x},\zeta)\psi(\varphi)=\psi' ,
\eqno (5.5)
$$
for all $\psi'$. With a standard notation, $\psi(\varphi)$ denotes
the boundary data, usually arranged in the form of a column vector
consisting of the field and its normal derivative(s) evaluated at
the boundary. The boundary conditions originally imposed read
(cf. (3.25) and (3.26), or (3.25) and (4.13))
$$
B \psi(\varphi)=0 \; \; {\rm at} \; \; {\partial M},
\eqno (5.6)
$$
where, in our case, the boundary operator takes the form
$$
B = \pmatrix{I & 0 \cr 0 & I \cr}.
\eqno (5.7)
$$
Unlike the case of mixed boundary conditions for the Laplacian, no
projectors occur in Eq. (5.7), because we are setting to zero on
the boundary the whole of $\varphi$ and its first (or second) normal
derivative. The graded leading symbol $\sigma_{g}(B)$ of $B$, is
defined according to the rule [10] 
$$
\sigma_{g}(B)_{ij} \equiv \sigma_{L}(B)_{ij} \; \; {\rm if}
\; \; {\rm ord}(B_{ij})=j-i,
\eqno (5.8a)
$$
$$
\sigma_{g}(B)_{ij} \equiv 0 \; \; {\rm if} \; \;
{\rm ord}(B_{ij})< j-i,
\eqno (5.8b)
$$
where ``ord'' denotes the order of the $B_{ij}$ component of $B$
as a differential operator on the boundary, subject to
the restriction
$$
{\rm ord}(B_{ij}) \leq j-i.
\eqno (5.9)
$$
By virtue of (5.7)--(5.9), the graded leading symbol of the
boundary operator is again the identity matrix
$$
\sigma_{g}(B)=\pmatrix{I & 0 \cr 0 & I \cr}
\eqno (5.10)
$$
in the boundary-value problems of Secs. III and IV, whereas
$\psi(\varphi)$ consists of the column vectors (here, the
symbol ${ }_{;N}$ coincides with $\nabla_{0}$ used in the
previous sections)
$$
\psi(\varphi)=\pmatrix{[\varphi]_{\partial M} \cr
[\varphi_{;N}]_{\partial M} \cr}
$$
or
$$
\psi(\varphi)=\pmatrix{[\varphi]_{\partial M} \cr 
[\varphi_{;NN}]_{\partial M} \cr}
$$
respectively, with $\psi'$ any pair of boundary data [9]
$$
\psi'=\pmatrix{\psi_{0}' \cr \psi_{1}' \cr}.
$$

After having defined in detail the strong ellipticity setting,
we can perform the next step, i.e. the solution of Eq. (5.3)
jointly with the conditions (5.4) and (5.5). Indeed, Eq. (5.3) reads
$$
\left[{d^{4}\over dr^{4}}-2|\zeta|^{2}{d^{2}\over dr^{2}}
+|\zeta|^{4} \right]\varphi(r)
=\lambda \varphi(r).
\eqno (5.11)
$$
Equation (5.11) is solved by
$\varphi(r)=e^{\alpha r}$, with $\alpha$ given by the roots
of the algebraic equation 
$$
\alpha^{4}-2\alpha^{2}|\zeta|^{2}+|\zeta|^{4}-\lambda=0,
\eqno (5.12)
$$
i.e.
$$
\alpha_{1}=+\sqrt{|\zeta|^{2}+\sqrt{\lambda}},
\eqno (5.13)
$$
$$
\alpha_{2}=+\sqrt{|\zeta|^{2}-\sqrt{\lambda}},
\eqno (5.14)
$$
$$
\alpha_{3}=-\sqrt{|\zeta|^{2}+\sqrt{\lambda}},
\eqno (5.15)
$$
$$
\alpha_{4}=-\sqrt{|\zeta|^{2}-\sqrt{\lambda}}.
\eqno (5.16)
$$
Comparison with the case of the Laplace operator [10] shows that one
obtains strong ellipticity provided that $\pm \sqrt{\lambda} \in
{\cal C}-{\Re}_{+}$, which yields
$$
\lambda \in ({\cal C}-{\Re}_{+}) \cap {\cal C}={\cal C}-{\Re}_{+}.
\eqno (5.17)
$$
Among the four values of $\alpha$, only $\alpha_{3}$ and $\alpha_{4}$
fulfill the condition (5.4). The exponentially decaying solution picked
out by (5.4) reads therefore ($\chi_{1}$ and $\chi_{2}$ being 
some parameters)
$$
\varphi(r)=\chi_{1}e^{-\rho_{1}r}+\chi_{2}e^{-\rho_{2}r},
\eqno (5.18)
$$
where 
$$
\rho_{1} \equiv +\sqrt{|\zeta|^{2}+\sqrt{\lambda}}, 
\eqno (5.19)
$$
$$
\rho_{2} \equiv +\sqrt{|\zeta|^{2}-\sqrt{\lambda}}. 
\eqno (5.20)
$$
One then finds
$$
\varphi|_{r=0}=\chi_{1}+\chi_{2},
\eqno (5.21)
$$
$$
\left . {d\varphi \over dr} \right |_{r=0}
=-\rho_{1}\chi_{1}-\rho_{2}\chi_{2},
\eqno (5.22)
$$
$$
\left . {d^{2}\varphi \over dr^{2}} \right |_{r=0}
=\rho_{1}^{2}\chi_{1}+\rho_{2}^{2}\chi_{2}.
\eqno (5.23)
$$
Thus, if the boundary conditions (3.25) and (3.26) are imposed,
the condition (5.5) leads to
$$
\pmatrix{I & 0 \cr 0 & I \cr}
\pmatrix{[\varphi]_{\partial M} 
\cr [\varphi_{;N}]_{\partial M} \cr}
=\pmatrix{\psi_{0}' \cr \psi_{1}' \cr},
\eqno (5.24)
$$
which implies, by virtue of (5.21) and (5.22),
$$
A \chi=\psi' ,
\eqno (5.25)
$$
where $\chi$ is the column vector $\pmatrix{\chi_{1} \cr 
\chi_{2} \cr}$, and $A$ is the matrix
$$
A \equiv \pmatrix{1 & 1 \cr -\rho_{1} & -\rho_{2} \cr}.
\eqno (5.26)
$$
Now a unique solution $\chi=A^{-1}\psi'$ exists of the linear
system (5.25), because
$$
{\rm det}(A)=\rho_{1}-\rho_{2} \not = 0.
\eqno (5.27)
$$
Explicitly, one finds
$$
\chi_{1}={(\psi_{1}'+\rho_{2}\psi_{0}')\over 
(\rho_{2}-\rho_{1})},
\eqno (5.28)
$$
$$
\chi_{2}=-{(\psi_{1}'+\rho_{1}\psi_{0}')\over
(\rho_{2}-\rho_{1})}.
\eqno (5.29)
$$
Moreover, if the boundary conditions (3.25) and (4.13) are imposed,
the condition (5.5) leads to
$$
\pmatrix{I & 0 \cr 0 & I \cr}
\pmatrix{[\varphi]_{\partial M} \cr 
[\varphi_{;NN}]_{\partial M} \cr}
=\pmatrix{\psi_{0}' \cr \psi_{2}' \cr},
\eqno (5.30)
$$
and this implies, by virtue of (5.21) and (5.23),
$$
{\widetilde A} \; \chi=\psi' ,
\eqno (5.31)
$$
where $\widetilde A$ is the matrix
$$
{\widetilde A} \equiv \pmatrix{1 & 1 \cr 
\rho_{1}^{2} & \rho_{2}^{2} \cr}.
\eqno (5.32)
$$
A unique solution $\chi={\widetilde A}^{-1} \; \psi'$ exists of the
linear system (5.31), because
$$
{\rm det}({\widetilde A})=\rho_{2}^{2}-\rho_{1}^{2}
=-2 \sqrt{\lambda} \not = 0.
\eqno (5.33)
$$
The explicit calculation yields (cf. (5.28) and (5.29))
$$
\chi_{1}={(\psi_{2}'-\rho_{2}^{2}\psi_{0}')\over
(\rho_{1}^{2}-\rho_{2}^{2})},
\eqno (5.34)
$$
$$
\chi_{2}=-{(\psi_{2}'-\rho_{1}^{2}\psi_{0}')\over
(\rho_{1}^{2}-\rho_{2}^{2})}.
\eqno (5.35)
$$

So far our examples of strong ellipticity have been almost 
straightforward. A more relevant case is obtained on considering
mixed boundary conditions for the squared Laplacian, with boundary
operator involving also tangential derivatives. For this purpose,
we assume that the boundary operator in (5.7) is replaced by
$$
B=\pmatrix{\Pi & 0 \cr \Lambda^{k} & I-\Pi \cr},
\eqno (5.36)
$$
where $k$ is an integer $\geq 1$, $\Lambda$ is a first-order
tangential differential operator on the boundary:
$$
\Lambda: C^{\infty}(W_{0},\partial M) \rightarrow
C^{\infty}(W_{0},\partial M),
$$
and $\Pi$ is a self-adjoint projector.
With a standard notation, $W$ is the bundle of boundary data over
$\partial M$, given by the direct sum
$$
W=W_{0} \oplus W_{1},
\eqno (5.37)
$$
with $W_{j}$ representing normal derivatives of order $j$. In the
previous case, we had $W$ expressed by (5.37) or, instead, by
$$
W=W_{0} \oplus W_{2}.
\eqno (5.38)
$$
The projector $\Pi$ maps each $W_{k}$ sub-space into itself.
The form of $\Lambda$ which leads to tangential derivatives in the
boundary conditions is
$$
\Lambda=(I-\Pi)\left[{1\over 2}\Bigr(\Gamma^{i}{\widehat \nabla}_{i}
+{\widehat \nabla}_{i}\Gamma^{i}\Bigr)+S \right](I-\Pi),
\eqno (5.39)
$$
where $\Gamma^{i}$ are endomorphism-valued vector fields on
$\partial M$, and $S$ is an endomorphism of the vector bundle
$W_{0}$. Following ref. 9, $\Gamma^{i}$ and $S$ are taken to be
anti-self-adjoint and self-adjoint, respectively, and annihilated
by the projector $\Pi$ from the left and from the right. By virtue
of the assumption (5.36), the graded leading symbol of the
boundary operator is (here $T \equiv \Gamma^{j}\zeta_{j}$)
$$
\sigma_{g}(B)=\pmatrix{\Pi & 0 \cr (iT)^{k} & I -\Pi \cr},
\eqno (5.40)
$$
and hence, when (5.37) holds, the boundary condition (5.5) 
takes the form
$$
\pmatrix{\Pi & 0 \cr (iT)^{k} & I-\Pi \cr}
\pmatrix{ \chi_{1}+\chi_{2} \cr -\rho_{1}\chi_{1}
-\rho_{2}\chi_{2} \cr}=\pmatrix{\psi_{0}' \cr 
\psi_{1}' \cr},
\eqno (5.41)
$$
where
$$
\psi_{0}'=\Pi(\gamma_{1}+\gamma_{2}),
\eqno (5.42)
$$
$$
\psi_{1}'=(I-\Pi)(-\rho_{1}\gamma_{1}-\rho_{2}\gamma_{2}).
\eqno (5.43)
$$
This means that the right-hand side of Eq. (5.41) is a smooth
section of an auxiliary vector bundle $W'$ over $\partial M$, where
any $\psi' \in C^{\infty}(W',\partial M)$ is given by 
$\psi'=P {\widetilde \psi}$, with
$$
P \equiv \pmatrix{\Pi & 0 \cr 0 & I-\Pi \cr},
\eqno (5.44)
$$
and ${\widetilde \psi} \in C^{\infty}(W,\partial M)$. The boundary
operator in (5.36) is related to the operator $P$ defined in (5.44)
by the equation
$$
B=P L,
\eqno (5.45)
$$
where
$$
L=\pmatrix{I & 0 \cr \Lambda^{k} & I \cr}.
\eqno (5.46)
$$
In other words, since we deal with a squared Laplacian, we allow 
for a number of tangential derivatives greater than the number
considered in the case of the Laplacian, and we try to impose 
mixed boundary conditions, so that projectors (rather than identity
operators) occur in the boundary operator. Since $\widetilde \psi$
differs in general from $\psi$, we have constants $\gamma_{1}$ and
$\gamma_{2}$ in (5.42) and (5.43) instead of $\chi_{1}$ and $\chi_{2}$
(cf. (5.18)).

Thus, for all $j=1,2$, if the matrix $(iT)^{k}-I \rho_{j}$ is
non-singular we may use the identity
$$
\chi_{j}=\Pi \chi_{j}+(I-\Pi)\chi_{j}
\eqno (5.47)
$$
to solve the system (5.41) in the form
$$
\Pi \chi_{j}=\Pi \gamma_{j},
\eqno (5.48)
$$
$$
(I-\Pi)\chi_{j}=\Bigr(I \rho_{j}-(iT)^{k} \Bigr)^{-1}
\left[(iT)^{k}\Pi +(I-\Pi)\rho_{j}\right]\gamma_{j}.
\eqno (5.49)
$$
A necessary and sufficient condition to obtain a unique solution
is of course the non-degeneracy of the matrix
$(I \rho_{j}-(iT)^{k})$, i.e.
$$
{\rm det} \Bigr[I \rho_{j}-(iT)^{k}\Bigr] \not = 0.
\eqno (5.50)
$$
Note now that (see (5.19) and (5.20))
$$
I \rho_{1}-(iT)^{k}=I \sqrt{|\zeta|^{2}+\sqrt{\lambda}}-(iT)^{k},
\eqno (5.51)
$$
$$
I \rho_{2}-(iT)^{k}=I \sqrt{|\zeta|^{2}-\sqrt{\lambda}}-(iT)^{k},
\eqno (5.52)
$$
and such matrices are never singular if the condition (5.17) is
satisfied and $iT$ is self-adjoint.
\vskip 0.3cm
\leftline {\bf 6. - Concluding remarks}
\vskip 0.3cm
Our paper has analyzed the squared Laplace operator 
$\cstok{\ }^{2}$ acting on symmetric rank-two tensor fields 
$\varphi_{ab}$ on (flat) Riemannian manifolds with boundary. Its 
original contributions, of technical nature, are as follows.
\vskip 0.3cm
\noindent
(i) Symmetry of $\cstok{\ }^{2}$ is achieved provided that both 
$\varphi_{ab}$ and its normal derivative 
$n^{p}\nabla_{p}\varphi_{ab}$, or $\varphi_{ab}$ and the second
normal derivative $n^{p}n^{q}\nabla_{p}\nabla_{q}\varphi_{ab}$, 
are set to zero at the boundary.
\vskip 0.3cm
\noindent
(ii) The resulting boundary-value problems are strongly elliptic
with respect to the cone ${\cal C}-{\Re}_{+}$.
\vskip 0.3cm
\noindent
(iii) Strong ellipticity with respect to ${\cal C}-{\Re}_{+}$,
in the case of mixed boundary conditions including
tangential derivatives, has also been proved. Interestingly, no
restriction involving $T$, and hence the matrices $\Gamma^{j}$, is
obtained unlike the case of an operator of Laplace type [9].
\vskip 0.3cm
Of course, in the case of symmetric rank-two tensor fields the
identity operator in Eq. (5.25) reads actually 
$\delta_{(a}^{c} \; \delta_{b)}^{d}$, but apart from such minor details,
all calculations in Sec. V prove indeed that the boundary-value 
problem $\Bigr(\cstok{\ }^{2},B \Bigr)$ is strongly elliptic with
respect to the cone ${\cal C}-{\Re}_{+}$, on considering the vector
bundle of symmetric rank-two tensor fields over $M$. Our proof is 
simple but of some interest, because it clearly shows the role played
by fourth-order operators in doubling the number of linearly
independent solutions of the eigenvalue equation (5.3) for the leading
symbol. Strong ellipticity is crucial to ensure the existence of the asymptotic
expansions used in the theory of heat-kernel asymptotics [10].
From the point of view of quantum field theory, this means that the
one-loop semiclassical approximation is well defined and can be
explicitly evaluated [19].

We find it appropriate to stress once more that the double integration
by parts used to derive Eq. (2.14) is necessary to recover the 
correct number (and form) of boundary conditions for a fourth-order
elliptic operator like $\cstok{\ }^{2}$. For example, its simplest
(but still useful) form, i.e. the operator $B \equiv {d^{4}\over dx^{4}}$
on a closed interval of the real line, satisfies the identity [19] 
$$ \eqalignno{
\; & (Bu,v)-(u,B^{\dagger}v) \cr
&=\left[{d^{3}u^{*}\over dx^{3}}v \right]_{0}^{1}
-\left[{d^{2}u^{*}\over dx^{2}}{dv\over dx}\right]_{0}^{1}
+\left[{du^{*}\over dx}{d^{2}v \over dx^{2}}\right]_{0}^{1}
-\left[u^{*}{d^{3}v \over dx^{3}}\right]_{0}^{1},
&(6.1)\cr}
$$
where $u$ is a vector in the domain of $B$, $v$ is a vector in the
domain of the adjoint $B^{\dagger}$ of $B$, and we use the 
definition of inner product [19]
$$
(u,v) \equiv \int_{0}^{1}u^{*}(x)v(x)dx.
\eqno (6.2)
$$
Domains of self adjointness of ${d^{4}\over dx^{4}}$ are therefore,
in particular, the set of functions belonging to
$AC^{4}[0,1]$ (this is the set of functions in $L^{2}[0,1]$ whose
weak derivatives up to third order are absolutely continuous in 
[0,1], which ensures that the weak derivatives, up to fourth order,
are Lebesgue summable in [0,1], and that all $u$ in the domain are
of class $C^{4}$ on [0,1]) and satisfying the boundary 
conditions [19] (cf. (3.25) and (3.26))
$$
u(0)=u(1)=0,
\eqno (6.3)
$$
$$
u'(0)=u'(1)=0,
\eqno (6.4)
$$
or Eq. (6.3) jointly with [19] (cf. (3.25) and (4.13))
$$
u''(0)=u''(1)=0.
\eqno (6.5)
$$
Thus, a complete correspondence can be established between the
boundary-value problem for the squared Laplace operator in one
dimension [19] and the more elaborated case studied in our paper.

It now appears both interesting and necessary to study a scheme 
more general than the one where $\eta_{ab},h_{ab}$ and their
first or second normal derivatives are set to zero at the boundary.
For this purpose, one may start again from Eq. (2.14), but requiring
that the integrand, as a whole, should vanish at $\partial M$. Such
an integrand $\sigma(\eta,h)$ may be re-expressed as
$$ \eqalignno{
\; & \sigma(\eta,h)=\Bigr[-K h_{ab}\nabla_{c}
+n^{p}(\nabla_{c}\nabla_{p}h_{ab})+K_{c}^{\; p}
(\nabla_{p}h_{ab})\Bigr]\nabla^{c}F^{ab}(\eta) \cr
&-\Bigr[-K \eta_{ab}\nabla_{c}
+n^{p}(\nabla_{c}\nabla_{p}\eta_{ab})
+K_{c}^{\; p}(\nabla_{p}\eta_{ab})\Bigr]
\nabla^{c}F^{ab}(h).
&(6.6)\cr}
$$
The non-trivial problem, however, is to understand how to derive
the boundary operator (5.36) from the analysis of (6.6). In the
previous section, the introduction of (5.36) was too simplified,
leaving aside the problem of integrating by parts in the action.
If it were possible to achieve this, it would then remain to be seen
whether such a kind of generalized boundary conditions for
$\cstok{\ }^{2}$ can be derived from an invariance principle, as is
indeed the case for the Laplacian itself [1,5,9], upon requiring
invariance under infinitesimal gauge 
transformations. Non-local boundary
conditions for $\cstok{\ }^{2}$ might also be studied with some profit,
following the recent attempts to consider a non-local formulation
of Euclidean quantum gravity based on integro-differential boundary
conditions [21]. Last, but not least, the resulting heat-kernel
asymptotics should be thoroughly developed, to supplement the recent,
encouraging progress in the case of generalized boundary-value 
problems for operators of Laplace type [22--25]. All this adds
evidence in favour of the problems of quantum field theory and 
spectral geometry being able to lead to a deeper vision in
modern mathematical physics [1].
\vskip 0.3cm
\centerline {${ }^{*}\; { }^{*}\; { }^{*}$}
\vskip 0.3cm
The author is much indebted to Ivan Avramidi for scientific 
collaboration on the techniques used in the present paper and for
comments which led to a substantial improvement.
This work has been partially supported by PRIN97 ``Sintesi''.
\vskip 0.3cm
\leftline {REFERENCES}
\vskip 0.3cm
\item {[1]}
ESPOSITO G., {\it Dirac Operators and Spectral Geometry}, in
{\it Cambridge Lecture Notes in Physics}, Vol. {\bf 12}
(Cambridge University Press, Cambridge) 1998.
\item {[2]}
ATIYAH M. F., PATODI V. K. and SINGER I. M., {\it Math. Proc.
Camb. Phil. Soc.}, {\bf 77} (1975) 43.
\item {[3]}
D'EATH P. D. and ESPOSITO G., {\it Phys. Rev. D}, {\bf 43}
(1991) 3234.
\item {[4]}
ESPOSITO G., {\it Qu\-an\-tum Gra\-vi\-ty, 
Qu\-an\-tum Cos\-mo\-lo\-gy a\-nd
Lo\-re\-nt\-zi\-an Ge\-o\-met\-ri\-es}, in {\it Lecture Notes
Phys., New Ser. m: Monographs}, Vol. {\bf m12}
(Springer--Verlag, Berlin) 1994.
\item {[5]}
ESPOSITO G., KAMENSHCHIK A. Yu. and POLLIFRONE G., {\it Euclidean
Quantum Gravity on Manifolds with Boundary}, in
{\it Fundamental Theories of Physics}, Vol. {\bf 85}
(Kluwer, Dordrecht) 1997.
\item {[6]}
DONALDSON S., {\it Bull. Am. Math. Soc.}, {\bf 33}
(1996) 45.
\item {[7]}
MORGAN J. W., {\it The Seiberg--Witten Equations and Applications 
to the Topology of Smooth Four-Manifolds} (Princeton University
Press, Princeton) 1996.
\item {[8]}
FRIEDMAN R. and MORGAN J. W., {\it Gauge Theory and the Topology of
Four-Manifolds} (American Mathematical Society) 1998.
\item {[9]}
AVRAMIDI I. G. and ESPOSITO G., {\it Commun. Math. Phys.},
{\bf 200} (1999) 495.
\item {[10]}
GILKEY P. B., {\it Invariance Theory, the Heat Equation, and the
Atiyah--Singer Index Theorem} (Chemical Rubber Company, Boca
Raton) 1995.
\item {[11]}
GRUBB G., {\it Func\-tio\-nal Cal\-cu\-lus 
of Pse\-u\-do Dif\-fe\-ren\-ti\-al
Bo\-un\-da\-ry Pro\-ble\-ms}, in {\it Pro\-gre\-ss 
in Ma\-the\-ma\-ti\-cs},
Vol. {\bf 65} (Birk\-h\"{a}u\-ser, Bos\-ton) 1996.
\item {[12]}
ERDMENGER J., {\it Class. Quantum Grav.}, {\bf 14} (1997) 2061.
\item {[13]}
ERDMENGER J. and OSBORN H., {\it Class. Quantum Grav.},
{\bf 15} (1998) 273.
\item {[14]}
BRANSON T. P., {\it Commun. Math. Phys.}, {\bf 178} (1996) 301.
\item {[15]}
AVRAMIDI I. G., {\it Phys. Lett. B}, {\bf 403} (1997) 280.
\item {[16]}
AVRAMIDI I. G., {\it J. Math. Phys.}, {\bf 39} (1998) 2889.
\item {[17]}
EASTWOOD M. and SINGER I. M., {\it Phys. Lett. A}, {\bf 107}
(1985) 73.
\item {[18]}
ESPOSITO G., {\it Phys. Rev. D}, {\bf 56} (1997) 2442.
\item {[19]}
ESPOSITO G. and KAMENSHCHIK A. Yu., {\it Class. Quantum Grav.},
{\bf 16} (1999) 1097.
\item {[20]}
AVRAMIDI I. G., ESPOSITO G. and KAMENSHCHIK A. Yu., 
{\it Class. Quantum Grav.}, {\bf 13} (1996) 2361.
\item {[21]}
ESPOSITO G., {\it Class. Quantum Grav.}, {\bf 16} (1999) 1113.
\item {[22]}
McAVITY D. M. and OSBORN H., {\it Class. Quantum Grav.},
{\bf 8} (1991) 1445.
\item {[23]}
DOWKER J. S. and KIRSTEN K., {\it Class. Quantum Grav.},
{\bf 14} (1997) L169.
\item {[24]}
AVRAMIDI I. G. and ESPOSITO G., {\it Class. Quantum Grav.},
{\bf 15} (1998) 281.
\item {[25]}
DOWKER J. S. and KIRSTEN K., {\it Class. Quantum Grav.},
{\bf 16} (1999) 1917.

\bye